\documentclass[sigconf]{acmart}
\AtBeginDocument{%
  \providecommand\BibTeX{{%
    \normalfont B\kern-0.5em{\scshape i\kern-0.25em b}\kern-0.8em\TeX}}}

\setcopyright{acmcopyright}
\copyrightyear{2018}
\acmYear{2018}
\acmDOI{XXXXXXX.XXXXXXX}

\acmConference[Conference acronym 'XX]{Make sure to enter the correct
  conference title from your rights confirmation emai}{June 03--05,
  2018}{Woodstock, NY}
\acmPrice{15.00}
\acmISBN{978-1-4503-XXXX-X/18/06}

\usepackage{enumitem}
\usepackage{multirow}
\usepackage{circledsteps}
\usepackage{algorithm}
\usepackage{algorithmic}
\usepackage[noabbrev]{cleveref}
\usepackage{subcaption}
\usepackage{newfloat}
\usepackage{listings}
\DeclareMathOperator*{\argmax}{arg\,max}

\definecolor{red}{rgb}{1, 0, 0}
\definecolor{blue}{rgb}{0, 0, 1}

\usepackage[inline]{./trackchanges}%
\copyrightyear{2023} 
\acmYear{2023} 
\setcopyright{rightsretained} 
\acmConference[AIES '23]{AAAI/ACM Conference on AI, Ethics, and Society}{August 8--10, 2023}{Montréal, QC, Canada}
\acmBooktitle{AAAI/ACM Conference on AI, Ethics, and Society (AIES '23), August 8--10, 2023, Montréal, QC, Canada}
\acmDOI{10.1145/3600211.3604655}
\acmISBN{979-8-4007-0231-0/23/08}

\begin{document}

\title[Measuring User-Perceived Value for Rejecting Machine Decisions]{How do you feel? Measuring User-Perceived Value for Rejecting Machine Decisions in Hate Speech Detection}

%Where There is Freedom of Speech There is Hate: Measuring User Perceived Value for Machine Decisions in Hate Speech Detection}


\author{Philippe Lammerts}
\email{philippelammerts@gmail.com}
\orcid{0000-0003-3164-5592}
\affiliation{%
  \institution{Delft University of Technology}
  \city{Delft}
  \country{The Netherlands}
}

\author{Philip Lippmann}
\email{p.lippmann@tudelft.nl}
\orcid{0000-0002-0139-1061}
\affiliation{%
  \institution{Delft University of Technology}
  \city{Delft}
  \country{The Netherlands}
}

\author{Yen-Chia Hsu}
\email{y.c.hsu@uva.nl}
\orcid{0000-0002-8016-7534
}
\affiliation{%
  \institution{University of Amsterdam}
  \city{Amsterdam}
  \country{The Netherlands}
}

\author{Fabio Casati}
\email{fabio.casati@servicenow.com}
\orcid{0000-0001-7591-9562}
\affiliation{%
  \institution{ServiceNow}
  \city{Santa Clara}
  \state{CA}
  \country{USA}
}

\author{Jie Yang}
\email{j.yang-3@tudelft.nl}
\orcid{0000-0002-0350-0313}
\affiliation{%
  \institution{Delft University of Technology}
  \city{Delft}
  \country{The Netherlands}
}


\begin{abstract}
Hate speech moderation remains a challenging task for social media platforms. % Manual moderation is infeasible and machine learning models for detecting hate speech are unreliable as they often perform poorly on unseen data. 
Human-AI collaborative systems offer the potential to combine the strengths of humans' reliability and the scalability of machine learning to tackle this issue effectively.  % A key component in such systems is decision handover from machines to humans. 
%
%Existing methods for task handover in human-AI collaboration only consider the decision confidence of machines while failing to take into account the cost of incorrect predictions. 
While methods for task handover in human-AI collaboration exist that consider the costs of incorrect predictions, insufficient attention has been paid to accurately estimating these costs.
% have focused on finding appropriate heuristics for task handover from machines to humans, usually according to machine confidence, but they often do not fully take. 
% In this work, we introduce into the hate speech detection task a human-AI collaboration framework that automatically rejects machine decisions for human moderation based on value assessment of machine decisions from a user-centered perspective.
In this work, we propose a value-sensitive rejection mechanism that automatically rejects machine decisions for human moderation based on users' value perceptions regarding machine decisions. 
% value assessment of machine decisions from a user-centered perspective.
% The rejector is value-sensitive due to the incorporation of how humans feel towards correct, incorrect, and rejected predictions regarding instances of hate speech into the formulation of the optimal threshold, thus allowing us to take the implications of the made decision into account. 
% To measure , we obtained the required data through a rigorous crowdsourced study 
We conduct a crowdsourced survey study with 160 participants to evaluate their perception of correct and incorrect machine decisions in the domain of hate speech detection, as well as occurrences where the system rejects making a prediction. Here, we introduce Magnitude Estimation, an unbounded scale, as the preferred method for measuring user (dis)agreement with machine decisions.
% on  to  users to rate different outcomes of hate speech moderation scenarios using Magnitude Estimation, allowing us to quantify their level of (dis)agreement with the choice made and subsequently determine the optimal threshold. 
Our results show that Magnitude Estimation can provide a reliable measurement of participants' perception of machine decisions. By integrating user-perceived value into human-AI collaboration, we further show that it can guide us in \emph{1)} determining when to accept or reject machine decisions to obtain the optimal total value a model can deliver and \emph{2)} selecting better classification models as compared to the more widely used target of model accuracy. %Our proposed value-sensitive rejection mechanism therefore, paves a way for bridging the gap between the current practice of human-AI collaboration and the goal of optimizing for user-centered value. 
% These results Our work paves the way for measuring and integrating into human-AI collaboration that can deliver higher 
\end{abstract}





\maketitle
 
\section{Introduction}
\label{s:1}

Hateful content spread online through social media remains a significant problem.
Ignoring its presence can lead to psychological harm and even result in violence and other conflicts~\cite{columbia-facebook-linked-to-violence, mujib-mashal-india, paul-mozur-2018, muller2021fanning}.
Governmental institutions and social media platforms are increasingly aware of these risks and are combating hate speech.
For example, the European Union developed a Code of Conduct on countering hate speech~\cite{eu-code-of-conduct}, requesting large social media companies to moderate hate speech and report their progress yearly. However, results reported so far are not yet satisfactory, as, for example, less than 5\% of hateful content has been removed from Facebook~\cite{noah2021giansiracusa}.

Hateful content moderation is either carried out manually or automatically by computational algorithms, where manual moderation may be more reliable but is not scalable to handle the deluge of user-generated content ~\cite{klonick2017new}. Further, continuous exposure to harmful content can be harmful to moderators as it can induce mental issues and potentially even lead to acts of self-harm~\cite{solonFacebook}. Computational solutions are, therefore, urgently in demand by online platforms~\cite{fortuna2018survey}. The methods considered best suited to this task are mainly based on machine learning, which has achieved reasonable performance at scale~\cite{gagliardone2015countering}. Yet, machine learning methods are far from being reliable, especially in dealing with hateful content previously unseen in the training data, which is often limited in size and biased~\cite{balayn2021automatic}. Several recent studies on hate speech have shown a significant drop in machine learning performance when assessed on different data from those captured in the training phase~\cite{arango2019hate,grondahl2018all}.

An approach that can combine the strengths of both previously mentioned approaches is human-AI collaboration, where humans are involved to solve AI-hard tasks, typically by taking over decisions where machines are unreliable~\cite{cheng2015flock,Law_hearth_cscw18}.
Such an approach is favorable in applications where decisions involve high-stakes and incorrect decisions can lead to damaging effects, as is the case for hate speech detection. 
Human-AI collaboration has been advocated in the human computation community~\cite{cheng2015flock,Raghu2019,WilderHK20} and, likely, is also an approach widely being used in enterprise applications such as search and conversational agents~\cite{Khodabandeh2020ExpandingAI}. Despite this, methods for implementing human-AI collaboration so far are limited to predefined heuristics and have largely ignored the complexity of real-world problems, especially the cost of incorrect predictions being context-dependent.

Common heuristics of task handover from machines to humans are based on machine confidence: humans take over the task when the confidence of the machine in its decision is lower than a predefined threshold \cite{Law_hearth_cscw18}. Such heuristics assume that machine confidence is well-calibrated, that is, a decision with high confidence should be more likely to be reliable and vice versa. This assumption however does not hold for many machine learning models, especially deep learning models, which may indicate high confidence when decisions are incorrect or vice versa~\cite{Guo2017,Balda2020}. An improved approach is proposed by~\citet{geifman2017selective} which determines the appropriate confidence threshold based on empirical evidence of machine correctness, e.g., based on the accuracy-threshold curve obtained on an empirical dataset. Such an approach, however, does not take into account the implications of right or wrong decisions. Incorrect decisions in high-stakes domains have a larger impact that, in turn, should pose a stricter constraint on accepting machine decisions, e.g., via a higher confidence threshold. Similar ideas have recently been discussed in position papers that advocate the adoption of the notion of context-dependent \emph{value} as a replacement of accuracy, the most common metric in machine decisions assessment~\cite{Sayin2021reject,casati2021value}. Value, however, is an abstract term -- it can be interpreted from social, ethical, or commercial perspectives~\cite{cummings2006ethics,zhu2018algo, gilliland2020value} -- yet the discussion on what creates value and how to measure it, specifically in a machine learning context, is limited due to it depending on the application. %In specific contexts, the question remains how value should be operationalized and embedded into the human-AI collaboration framework. 

In this paper, we study the problem of operationalizing value perception of machine decisions and its integration into human-AI collaboration in the specific context of hate speech detection. % \rmj{We first introduce a formal framework of human-AI collaboration for the chosen task where value is integrated as an abstract parameter that determines the optimal confidence threshold for rejecting machine decisions (and subsequently passing the decision to human moderators).} 
We start by identifying several factors that may affect the value definition, %\rmj{ used for such a framework}, 
namely the selection of a specific stakeholder's standpoint and the relativity of value perception as affected by stakeholder expectation or regulation. We then operationalize user-perceived value in hate speech moderation scenarios, where a decision with a corresponding confidence has been made by a machine. To measure these perceptions, we explore several measurement scales and propose to select Magnitude Estimation (ME)~\cite{stevens1956direct} as the primary scale. ME allows the measurement of the magnitude of user (dis)agreement using an unbounded scale and makes it possible to obtain the relative ratios between the magnitudes of different machine decisions. These ratios are essential to determine the optimal confidence threshold for rejecting machine decisions (see \cref{s:3}).

To validate ME in value operationalization, we designed a survey study where we recruited 160 participants. Each participant's perception regarding a dataset of 40 selected hateful and non-hateful tweets and their (dis)agreement regarding the corresponding machine decisions were evaluated. Through a between-subject study, we show that Magnitude Estimation returns results with significantly higher inter-rater reliability compared to other scales, showing its suitability in measuring user perception. Our results show that the inter-rater reliability is significantly higher for incorrect decisions than for correct decisions, indicating a strong consensus among participants regarding the consequences of harm, as well as disagreements on what constitutes hate online.
Further, users appear to be more negatively affected when a non-hateful post is subject to moderation than when an instance of hate speech is classified as non-hateful, implying that users would rather contend with an instance of hate speech than have an innocent user punished for a non-hateful post. 
%\jie{say also the higher agreement in value rating for incorrect decisions, and what it entails} \Philippe{done, correct?} % Our results also show that users tend to disagree more on false prediction for hateful content classified as non-hateful than the other way around. 

To demonstrate the utility of value integration in human-AI collaboration, we evaluate the effect of rejecting machine decisions made by three machine learning-based hate speech detection models -- including traditional, deep learning, and BERT-based models \cite{devlin2018bert} -- in handling data from both seen and unseen sources. Our results show that for all three models, when evaluated on unseen data, the optimal confidence thresholds determined by the model-delivered value are much higher than the optimal thresholds on seen data. These results confirm the findings from previous studies on machine biases and demonstrate the effectiveness of using value as a target for optimally rejecting machine decisions. We further show that when selecting the optimal model, using value as the criterion returns different results compared to using accuracy. Note, that our approach to measuring value perception can be applied to different tasks and is model-agnostic.

In summary, we make the following key contributions:
\begin{itemize}
    % \item \rmj{We introduce the human-AI collaboration framework with a rejection mechanism for machine decisions to the problem of hate speech detection;}
    \item We introduce Magnitude Estimation as a scale for measuring user perception of machine decisions in scenarios where these decisions are correct and incorrect;
    \item We demonstrate the applicability of Magnitude Estimation through a between-subject survey study, as well as the utility of value for optimally rejecting machine decisions;
    % \item As part of our survey study we find that users tend to be more in agreement regarding what constitutes a misclassified hate speech instance than what constitutes a correctly classified instance. Users also appear to be more negatively affected when a non-hateful post is subject to moderation than when an instance of hate speech is classified as non-hateful.
    \item We contribute a set of insights into user-perceived value of automated machine decisions, especially their  attitudes towards different types of (mis)classifications.
\end{itemize}

\section{Background on Value-Sensitive Rejection of Machine Decisions}
\label{s:3}
This section introduces the background of value-sensitive rejection of machine decisions in a hybrid human-AI workflow, based on previous work \cite{Sayin2022,Sayin2021reject}, and subsequently identifies factors that influence value perception in hate speech detection.

\subsection{Rejection for Binary Classification} 
We consider the general case of human-AI collaboration as follows: the machine decision can either be accepted or rejected; if rejected, the decision will be taken over by a human decision maker. 
Formally, consider a binary classification problem for which we have a machine learning classifier, whose output on a data item $x$ is confidence, $\mathbf{c}$, (e.g., the output from the softmax layer of a neural network). The rejection is dependent on a threshold denoted by $\tau \in [0,1]$, which then modifies the final output of the machine as
\begin{equation}
    \hat{y} = 
    \begin{cases}
       \text{$y$,\ } \quad\text{$\mathbf{c}_y \geq \tau$,}\\
       \text{$y_r$,} \quad\text{otherwise.} \\ 
     \end{cases}
\end{equation}
where $y$ denotes an accepted decision and $y_r$ denotes the special decision of rejection, resulting in a human making the final decision. 

We now discuss how the optimal confidence threshold for rejecting machine decisions is affected by the value formulation. 
%
%Without loss of generality, 
We consider the binary classification case: when the machine decision is either positive (i.e., the content is deemed hateful) or negative (i.e., non-hateful). There is a value, $V$, attached to each of these, depending on whether this positive or negative decision is correct or not. This results in true positive (TP), true negative (TN), false positive (FP), false negative (FN), and rejected predictions as possible outcomes. $V_{TP}$ and $V_{TN}$ are positive, while $V_{FP}$, $V_{FP}$, and rejected predictions, $V_r$, are negative (i.e., costs).
%We further consider the case when the classifier is well-calibrated, as well as when the classifier is not calibrated. 
% \subsubsection{Optimal Threshold for Calibrated Machines}
\iffalse
Consider that for the positive classifications, the accuracy of the classifier is $\alpha^p$, meaning that the $\alpha^p$ proportion of the positive classifications are true positives, and $(1-\alpha^p)$ are false positives. \color{red}Given this, the optimal threshold $\tau^p$ should be the one where the total value is greater than 0 when the confidence of machine decision is above $\tau^p$ and less than 0 otherwise. Equivalently, it means the value of acceptance of the machine decision when the confidence is $\tau^p$ should be 0. When the model is calibrated, we have $\alpha^p = \tau^p$; that is, the model confidence equals the accuracy, resulting in\color{black}
\begin{equation}
\tau^p V_{TP} + (1-\tau^p) V_{FP} = 0
\end{equation}
where $V_{TP}$ and $V_{FP}$ are the value of true positive and false positive outcomes\cmj{let's add a footnote here that the value of false positive $V_{FP}$ should be a negative number, indicating essentially a cost}, respectively. 
\fi
The optimal threshold for positive classifications is:
\begin{equation}
    \tau^p_O = \frac{V_{FP}}{V_{FP} - V_{TP}} = \frac{\gamma^p}{\gamma^p+1}
    \label{eq:optimal_threshold}
\end{equation}
if we assume $V_{FP} = -\gamma^p \cdot V_{TP}$, that is, the cost of a false positive is $\gamma^p$ times worse than the value of a true positive. Similarly, in the case of negative classifications, the optimal threshold would be $\tau_O^n = \frac{\gamma^n}{\gamma^n+1}$ where $V_{FN} = -\gamma^n \cdot V_{TN}$, i.e., the cost of false negative is $\gamma^n$ times worse than the value of a true negative.

When the cost of incorrect decisions is very high, i.e., $\gamma \gg 1$, the optimal confidence threshold would tend close to 1, meaning almost all machine decisions are rejected. When the cost of an incorrect decision is very low, i.e., $\gamma \approx 0$, the optimal threshold would be close to 0, and virtually all machine decisions are accepted. These results, therefore, follow our intuition. 
An important conclusion we can draw from~\cref{eq:optimal_threshold} is that the optimal threshold is dependent \emph{only on the ratio} of the value (or cost) between an incorrect decision and that of a correct one (per class). % instead of the absolute quantities. 

% \subsubsection{Optimal Threshold for Non-Calibrated Machines} 
 
%As different thresholds lead to different total values, the optimal threshold that corresponds to the highest overall value can be selected.
%
Threshold optimization is the process of finding the threshold that maximizes value empirically. If a system is calibrated before use, simulations can be used to find the optimal theoretical threshold, which is the optimal $\tau$ that maximizes value. In this paper, $\tau$ is determined by means of calibration, done by means of temperature scaling~\cite{Mozafari2018scaling}, followed by a calculation of the theoretical threshold based on the crowdsourced survey data, as it allows us to quantify and compare the opinions of participants on the value of true and false predictions and thus compute the ratios for our use case.

\subsection{Value Factors in Hate Speech Detection}
%\subsubsection{Value Characteristics in Hate Speech Detection}
We denote the value of classifying a data item correctly, or incorrectly, and that of rejecting a classification as ${V}_{c}$, ${V}_{w}$, and $V_r$, respectively. We make the following observations when considering value for hate speech detection: 1) Value is dependent not only on the machine learning model but also on the specific context to which the model is applied. For example, an incorrect prediction in the medical domain potentially has a bigger impact than one in e-commerce. In a high-stakes domain, generally, we would assume ${V}_{c} > V_r > {V}_{w}$ and thus a correct machine decision saves the cost of human moderation and accelerates the decision-making process, while a rejection requires additional human intervention. 2) Value interpretations from different stakeholders can vary. In hate speech detection, for example, a rejection of a machine decision induces the cost of human moderation from the business perspective, while from the user perspective what is more important is the exposure to hateful content. In our study, we choose to take the user's standpoint, and, as such, view $V_r$ to come with an inherent cost since human moderation will be pending and the potentially hateful content will remain visible. 3) Value is affected by both stakeholder expectations and regulation.
For example, in the hate speech detection case, when hateful content is posted, from the user's perspective, the value derived from a correct machine decision depends on the user's general expectation of how hateful content should be handled. Similarly, the legality of hate speech in certain jurisdictions may influence stakeholder perception.
%
%If the expectation is that generally, the platform takes no action, then a correct machine decision would be highly appreciated, hence it comes with a high value; otherwise 
%If users generally expect that hateful content can be detected and removed by the platform, then a correct machine decision has less value -- though the value should still be higher than zero since automation accelerates decision making.%, compared to the baseline of human moderation that takes time. 
%

%Regarding regulation, when legal requirements are in place, the value of correct machine decisions will be less appreciated while incorrect ones may be penalized. 

%\subsubsection{Calculation of Total Value}
Given the above observations, we now introduce the function to determine the total value, $V(\tau)$, of a given model with a reject option at the rejection threshold $\tau$ on a given dataset. Assuming that when accepted, correct decisions increase the overall value and when rejected, they decrease the overall value and vice versa, then, $V(\tau)$ may be formalized as: %We extended the work of~\citet{de2000reject} by introducing a $V(\tau)$ that measures the total value of the model and is based on the crowdsourced values of TP, TN, FP, FN, and rejected predictions.
\begin{align}
    \label{for:final-V}
    V(\tau) = \sum_{p} (V_p - V_r)N_p + \sum_{q} (V_r - V_q)N_{q},
\end{align}
where $p \in [TP, TN, FP, FN]$, $q \in [TP, TN, FP, FN]$, and $N_p$ and $N_q$ are the number of accepted and rejected data items for the difference scenarios, respectively. Note, that we assume that rejected decisions have a cost that decreases the overall value, i.e., $V_r$ is negative, as users have to wait on a moderation decision. Thus,~\cref{for:final-V} allows us to summarize the value gained and the cost subtracted into a single value for the model by considering the value or cost of each scenario and how often it occurs, while also taking the cost of rejection into account.

\section{Survey Study}
\label{sec:survey_design}

To define the relative value of scenarios, we design a survey to ask participants the degree to which they agree or disagree with the decisions of a fictional social media platform, SocialNet.
These scenarios represent TP, TN, FP, FN, and rejected predictions.
The TP and TN scenarios imply that SocialNet successfully detects whether a post is hateful or not hateful, respectively.
The FP scenario means that SocialNet incorrectly predicts a non-hateful post as hateful, and conversely for the FN scenario.
For example, in the FN scenario, the survey shows a hateful post to the subject and explains that SocialNet did not identify the post as hate speech.
%
%Then, participants indicate the degree of agreement/disagreement using some scale, and the answers per scenario are aggregated to obtain the relative values.
%
%For example, the FN scenario is considered $q$ times as worse as an FP scenario, where $q$ is computed based on the scale.

\subsection{Choice of the Scale}

We use ME as the primary scale.
A Likert scale was initially considered, as it is widely used in research for retrieving participant opinions and is perhaps more intuitive for participants~\cite{boone2012analyzing}.
%
%For example, we may use seven Likert-scale items ranging from ``strongly disagree'' to ``strongly agree'' with a ``neutral'' midpoint.
%
However, a Likert scale is not suitable in our case, as Likert-type items are ordinal, meaning that we only know the ranks but not the exact distances between the items~\cite{allen2007likert}.
In our case, computing the relative values (i.e., ratios) of our scenarios requires measuring the distances between different items, which cannot be provided by a Likert scale.
On the contrary, the ME scale allows us to measure ratios by asking participants to provide numerical ratings.
ME originated from psychophysics, where participants gave quantitative estimates of sensory magnitudes~\cite{stevens1956direct}.
For sound loudness, a sound twice as loud as the previous one, should ideally receive a rating twice as large.

Researchers have previously applied the ME scale to different physical stimuli (e.g. line length, brightness, or duration) and proved that the results are reproducible, as well as that the data has ratio properties~\cite{moskowitz1977magnitude}.
Other works have shown that the ME technique is also helpful for rating abstract types of stimuli, such as judging the relevance of documents~\cite{maddalena2017crowdsourcing}, the linguistic acceptability of sentences~\cite{bard1996magnitude}, and the usability of system interfaces~\cite{mcgee2004master}.
Thus, we conclude that ME is a promising method for judging hate speech.

\subsection{Normalization and Validation of the Scale}

The ME scale is unbounded.
For example, suppose we first show a scenario and the participant provides a value (e.g., 100) to indicate the degree of agreement.
Suppose we next present a scenario that the participant agrees with more.
The participant can always provide a higher value (e.g., 125) and not be restricted within a fixed range. %(e.g., between 0 and 100). 
%
%However, there are drawbacks to using the ME scale.
%
The results need to be normalized as different participants rate the agreement/disagreement degree differently.
%
%And it can be hard to validate if the ME scale measures the participants' judgments for different scenarios of hate speech detection consistently.

Multiple solutions exist for normalizing the ME scale, such as modulus normalization, which uses geometric averaging to preserve the ratio information~\cite{moskowitz1977magnitude, mcgee2004master}.
Unlike the unipolar ME scales used in previous research~\cite{bard1996magnitude, mcgee2004master}, we use bipolar scales.
% (disagree-agree) with 0 (neutral) and negative values (disagree).
%
%Thus, we cannot use geometric averaging since it uses logarithm calculations.
%
Using arithmetic averaging is inappropriate since it uses logarithmic calculations and would disrupt the ratio scale properties~\cite{moskowitz1977magnitude}.
%
% Another normalization approach is external calibration, which keeps the ratio scale properties~\cite{moskowitz1977magnitude}.
%
%This approach asks participants at the end of the survey to indicate which verbal label (e.g., ``strongly agree'') corresponds to which numerical value (e.g., 125) they used before.
%
Therefore, we normalize the results by dividing the magnitude estimates of each subject by their maximum estimate.
We multiply the normalized magnitude estimates by 100 for the sake of clarity.
This way, all magnitudes estimates are in the range $[-100, 100]$ while maintaining their ratio properties.
% Then, the average value is calculated as the pivot, and all the ME results are divided by the pivot value.
%
% In this way, external calibration normalizes ME results while maintaining the ratio properties.

%\subsection{Validation of the Scale}

Most previous research using the ME scale applies validation, such as cross-modality validation, where estimated magnitudes are compared to the physical stimuli using correlation analysis~\cite{bard1996magnitude}.
%
%For example, when asking a participant to estimate line lengths in two scenarios, we can use two lines where one's length is $r$ times longer than the other.
%
%Then, we can vary $r$ for each participant to compare the relationship between the ME results and the exact line lengths.
Cross-modality validation is difficult in domains that do not have exact measures of stimuli, such as hate speech.
%
%One alternative is to compare different non-physical stimuli, which has been proven suitable to validate the ME scale~\cite{mcgee2004master, lodge1979comparisons}.
%
%Some previous work analyzed the correlation between different ME scales~\cite{bard1996magnitude, lodge1976calibration}. %, such as handgrip measurements or drawing lines \cite{bard1996magnitude, lodge1976calibration}.
%
Some previous work compared ME with other validated scales~\cite{maddalena2017crowdsourcing}.
%that can be of any type
%
In our case, we use the 100-level scale to validate the ME scale by analyzing their correlation~\cite{roitero2018fine}, which is a form of convergent validation~\cite{fitzner2007reliability}.

\subsection{Participants and Data}
We use Prolific to recruit crowd workers for the study.\footnote{Approved by the ethics committee of our organization.}
Participants need to be at least 18 years of age, be fluent in English, and have an approval rating of over 90\%.
Participants also need to have experience using a social media platform regularly (at least once a month). %one of the following social media platforms regularly (at least once a month): Facebook, Twitter, YouTube, LinkedIn, Pinterest, Google Plus, Tumblr, Instagram, Reddit, VK, Flickr, Vine.co, Meetup, ask.fm, Snapchat, TikTok, or Medium.
Every participant is paid an hourly wage of 9 GBP, exceeding the UK minimum wage at the time of the study.
Regarding sample size, we recruit 24 participants for the pilot study and 136 participants for the official study.
Of the recruited participants, 50\% identified as female, though \citet{gold2018women} showed that there is no significant difference when perceiving hate between genders.
Half of the participants are assigned the ME scale and the other half the 100-level scale.
We choose a 90\% Confidence Interval (CI) and 10\% Margin of Error (MoE) for this study due to budget limitations.
There are billions of social media users, and according to~\citet{muller2014survey}, we need a sample size of 68 participants per measurement scale, i.e., 136 participants, to reach the desired CI and MoE.

%\subsection{Hate Speech Data}
The final dataset consists of 20 hateful and 20 non-hateful social media posts from a public dataset~\cite{basile2019semeval} to build the machine decision scenarios (TP, TN, FP, FN, and rejection).
The dataset contains 13,000 English tweets, and each tweet is annotated with three categories: hate speech (yes/no), target (group/individual), and aggressiveness (yes/no).
We first exclude tweets that are replies or contained mentions or URLs since they have unclear contexts.
Finally, we use clustering analysis to select 40 tweets for our study.
We use a cluster size of 20 for the non-hateful tweets and sample one tweet per cluster by taking the nearest sample to each cluster centroid to obtain each cluster's most representative tweets.
For the hateful tweets, we first divide them into four groups using the target and aggressiveness categories.
Similarly, for each hateful tweet group, we use a cluster size of 5 and sample one tweet per cluster.
We perform latent semantic analysis (LSA), which is a combination of term frequency-inverse document frequency (TF-IDF) and Singular Value Decomposition (SVD), and k-means clustering on each group of tweets.
We calculate the silhouette coefficient to determine the optimal cluster size ($k$ value) for the neutral tweets and the four groups of hateful tweets.
%
%\color{red}The silhouette analysis indicated to set $k$ as large as possible. \color{black}
%
We manually select one tweet per cluster using a majority vote from three members of our group to choose representative tweets and create the final set of 40 tweets.

Additional information on the study's variables, pilot study, demographics, as well as example tasks may be found in~\cref{app:b}.

\subsection{Procedure and Data Quality Control}
The survey first presents the informed consent policy and excludes participants that do not agree with it.
Next, introductory texts are shown to explain the possible machine decisions.
In the case of using the ME scale, participants are presented with a warm-up task to estimate different line lengths.
Then, the survey asks 40 randomly shuffled question sets regarding the TP, TN, FP, FN, and rejection scenarios (with 8 question sets per scenario).
%
%Each set contains several questions with the same structure.
%
The first question is about whether participants think the post is hateful (yes/no).
The second question is whether participants agree or disagree with the decision made by the machine, which may be correct or incorrect, or are neutral towards it.
In the case of a non-neutral decision, the survey asks the third question about the degree to which participants agree or disagree with the machine's decisions, using either the ME or 100-level scale, depending on their group.
There is no time limit for the survey.

%\subsection{Data Quality Control}

In the middle of the question sets, we use two Instructional Manipulation Checks to determine if the user is paying attention\footnote{\href{https://researcher-help.prolific.co/hc/en-gb/articles/360009223553}{Prolific's Attention and Comprehension Check Policy}}.
These attention checks ask participants to select a specific option from multiple choices (e.g., "You must select Orange").
We exclude responses from the participants who fail the attention checks or do not complete all questions.
For the ME scale, we discard responses that do not perform well in the line length warm-up task.

\iffalse
\subsection{Pilot Study}
\color{blue}
We conducted a pilot study with 24 participants to test the survey and estimate the required completion time.
%
The pilot study showed low inter-rater reliability, which may be due to unmatched expectations and unclear scenario descriptions.
%
Initially, our scenarios mentioned that machine-detected hateful posts would be removed, which could be controversial.
%
Thus, we changed scenarios to rank the hateful posts lower on the feed. % so that it takes more effort for users to scroll the page to access the posts.
%
Moreover, we added more lengthy descriptions to explain the potential consequences of all scenarios in a neutral tone, which prepares the participants to focus on evaluating harm (instead of giving rewards).
%
%After the changes, the official survey study shows higher inter-rater reliability.
\color{black}\fi

\subsection{Analysis}
\label{sec:analysis}

We first compute the values for the TP, TN, FP, FN, and rejection scenarios using the survey study data.
For both scales, we convert disagreement (with the machine decision) ratings to negative values, neutral stances to 0, and agreement ratings to positive values.
%
%Then, the ME values are used to compute the optimal rejection threshold, as described in section~\ref{s:3}).
%
%Next, we use the 100-level values to validate the ME values. 
%
We apply convergent validity, in which a correlation analysis between different scales (i.e., the ME and 100-level scales) is conducted to determine if they measure the same phenomenon \cite{fitzner2007reliability}.
We expect a medium-large correlation between both scales, meaning that ME responses small in magnitude should correspond to 100-level scale responses small in magnitude and vice versa.
Finally, we analyze reliability, which determines whether we can trust our results and achieve consistent outcomes~\cite{fitzner2007reliability}.
In our case, we use inter-rater reliability to investigate whether different subjects give approximately the same judgments to the same scenarios and, thus, whether the degree to which hate speech is subjective.
It is measured using Krippendorff's alpha, which we calculate using the normalized ME and 100-level values for all scenarios.
\section{Results}
\label{s:5}

%This section shows the survey results of value ratios, reliability, and validity (defined in~\cref{sec:survey_design}).
%
%Then, we present the value of different models at different rejection thresholds.\footnote{All our code and data are submitted in the supplementary files.}
%As explained in~\cref{sec:survey_design}, we will evaluate the survey results by focusing on three different aspects: value ratios, reliability, and validity, before showing the total value of different models at all thresholds.
\subsection{Reliability and Validity}
%\subsection{Value Ratios}
\label{sec:results-costs}

First, for each survey question set, we calculate the median of all responses.
This step yields 40 values (eight values per scenario).
We use the median since data from both scales are highly skewed.
Then, we calculate the mean of the values ($V_{TP}, V_{TN}, V_{FP}, V_{FN}, V_r$) within each scenario, giving us the final five values for the TP, TN, FP, FN, and rejection cases.
The results for both scales can be seen in~\cref{tab:costs-reliability}.
The total value, $V$, is calculated at a later point in this section using the different values.
\begin{table}[t]
\centering
\begin{tabular}{lcccc}
\toprule
& \multicolumn{2}{c}{\textbf{ME}} & \multicolumn{2}{c}{\textbf{S100}}\\
\cmidrule(l){2-3} \cmidrule(l){4-5}
& $\boldsymbol{\alpha}$ & $\textbf{v}$ & $\boldsymbol{\alpha}$  & $\textbf{v}$ \\
\midrule
\textbf{TP}          & 0.07    & 18.15  & 0.04 & 77.00                                \\
\textbf{TN}          & 0.10    & 36.32  & 0.11  & 86.31                               \\
\textbf{FP}          & 0.39    & -16.69 & 0.07  & -51.00                               \\
\textbf{FN}          & 0.92    & -28.08 & 0.14  & -62.43                               \\
\textbf{Rejection}   & -0.31   & -4.82  & 0.07  & -16.37                               \\
\midrule
\textbf{All}         & 0.78    & ---       & 0.44                               &    ---                            \\
\bottomrule
\end{tabular}
\caption{Krippendorff's alpha ($\alpha$) and the scenario values ($v$) for TP, TN, FP, FN, and rejection scenarios. ME refers to Magnitude Estimation, and S100 refers to the 100-level scale.}
\label{tab:costs-reliability}
\end{table}

%\subsection{Reliability and Validity}
We calculate Krippendorff's alpha to measure the inter-rater reliability of all scenarios for each scale, as shown in~\cref{tab:costs-reliability}.
The last row of the table contains the $\alpha$ values for the entire scale, measuring the inter-rater reliability for all answers.
We observe that the ME scale has high inter-rater reliability while the 100-level scale is less reliable.
Also, participants using the ME scale tend to exhibit higher agreement regarding the FP and FN cases and systematically disagree on the rejected cases.
For the 100-level scale, we observe that participants have low agreement on all scenarios. 

%\subsection{Validity}

\begin{figure}[h]
    \centering
    \hspace*{-0.3in}
    \includegraphics[width=0.8\linewidth]{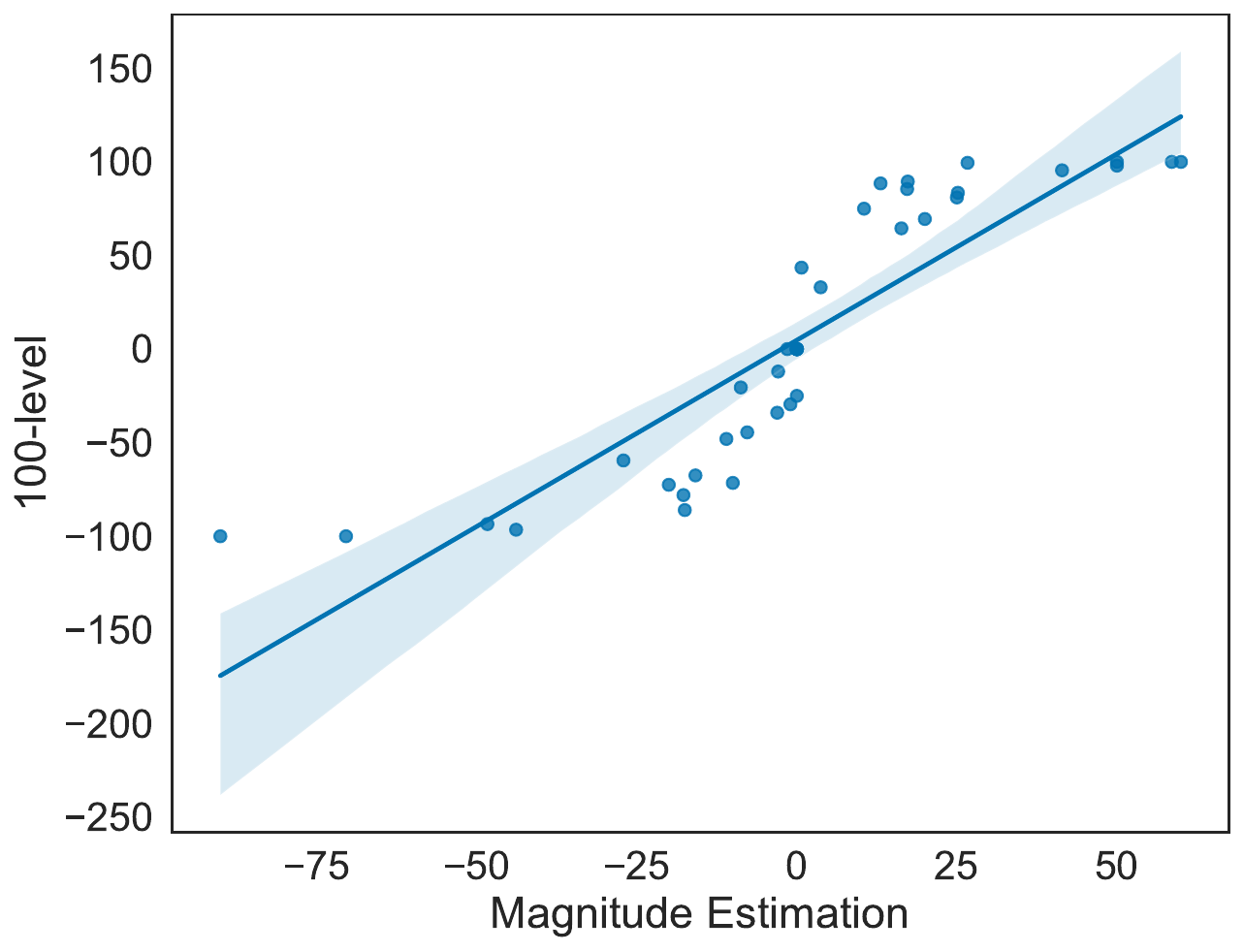}
    \caption{Correlation plot between the median normalized magnitude estimates and the median 100-level scores per question, showing agreement and disagreement.}
    \label{fig:validity}
\end{figure}

We analyze the validity of the ME scale by comparing the median normalized magnitude estimates with the median 100-level scores for each question set.
Figure~\ref{fig:validity} presents the correlation plot between the two scales.
A Shapiro-Wilk test indicates that the data of both scales do not follow a normal distribution ($p<0.05$).
Thus, we use the Spearman and Kendall rank correlation coefficients since these are non-parametric tests.
Spearman returned a $0.98$ and Kendall a $0.89$ correlation between the ME and the 100-level scales ($p<0.05$).
Finally, a Mann-Whitney U test between the ME and 100-level scales gives a large p-value, indicating no statistically significant difference between the two scales.
%shows that there the distributions of both scales are identical.

\subsection{Total Model Value due to Threshold}
\begin{figure}[h]
\centering
    \setkeys{Gin}{width=\linewidth} %set image parameters
    \begin{subfigure}[b]{.49\linewidth}
        \centering
        \includegraphics{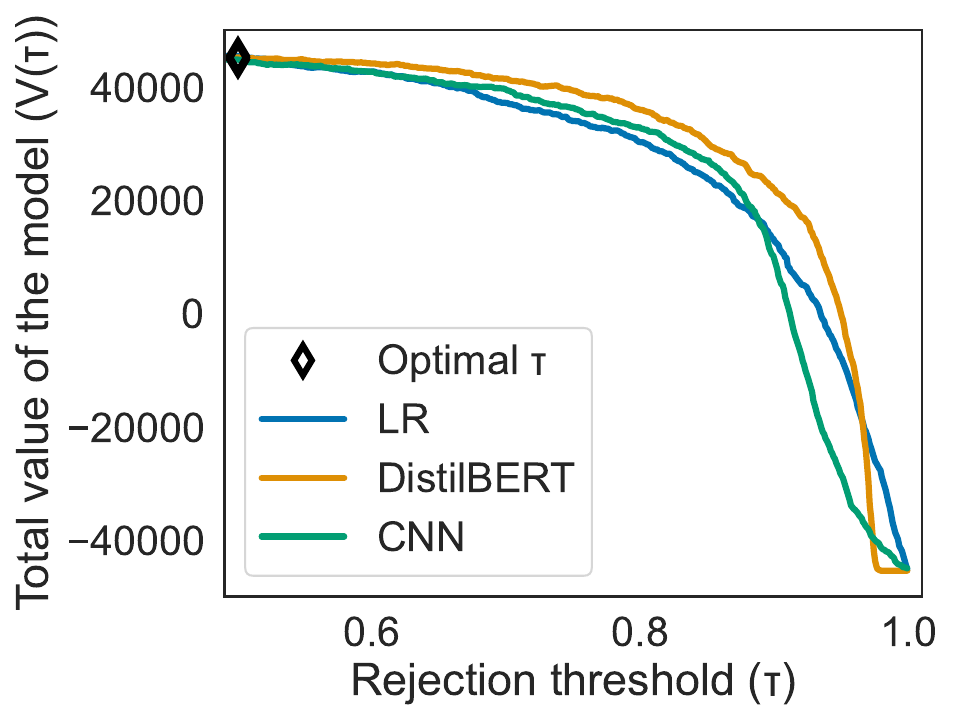}
        \caption{evaluated on \textit{seen} data}
        \label{fig:metric-values-seen}
    \end{subfigure}
    \hfill
    \begin{subfigure}[b]{.49\linewidth}
        \centering
        \includegraphics{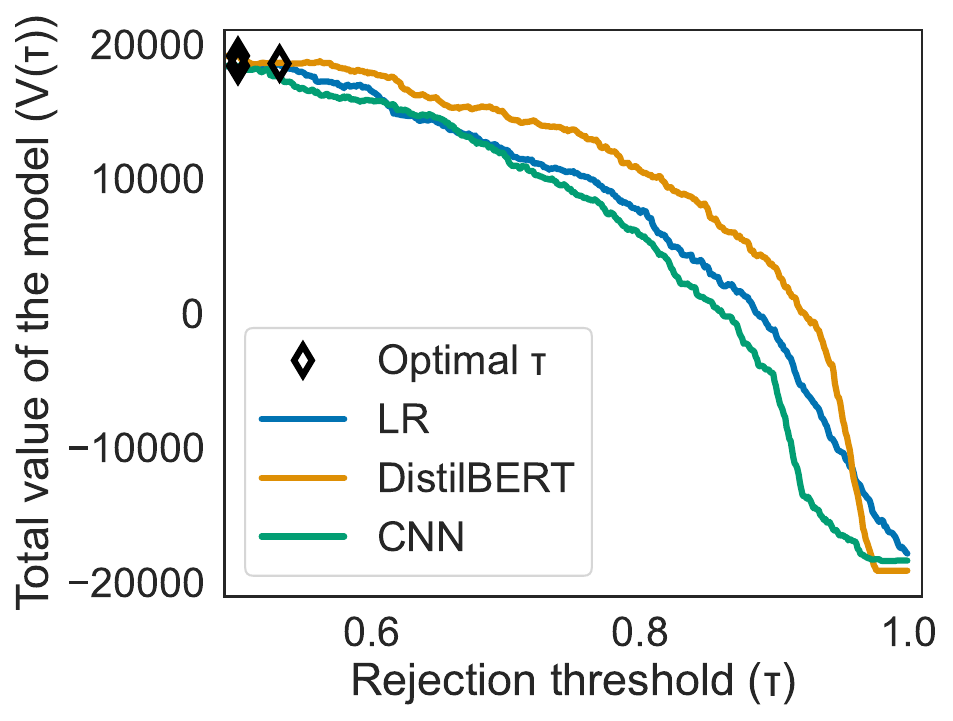}
        \caption{evaluated on \textit{unseen} data}
        \label{fig:metric-values-unseen}
    \end{subfigure}
    \label{fig:metric-original-values} % does not work, gets overwritten below
    \caption{$V(\tau)$ curves of all models with $v$ of TP=18.15, TN=36.32, FP=-16.69, FN=-28.08, and rejection=4.82.}\smallskip
\end{figure}

We evaluate the $V(\tau)$ function (i.e., the value at different rejection thresholds) using the values from the survey study obtained using the ME scale.
We train three different binary hate speech classification models on the~\citet{waseem2016hateful} dataset.
The used models are Logistic Regression (LR) with Character N-gram~\cite{waseem2016hateful}, a Convolutional Neural Network (CNN) based on~\citet{agrawal2018deep}, and a DistilBERT transformer~\cite{sanh2019distilbert}.
We use Temperature Scaling to calibrate the CNN and the DistilBERT models following the approach from~\citet{Guo2017}.
The model predictions are based on two different test datasets: the \textit{seen} dataset and the \textit{unseen} dataset.
The \textit{seen} dataset is the test set of~\citet{waseem2016hateful} and the \textit{unseen} dataset is a test set from a separate but similar source~\cite{basile2019semeval}.
We use the \textit{unseen} dataset to simulate how the models would perform in a more challenging, realistic use case.
Using unseen data that is similar but separate from the training set, we also investigate the impact of bias.
Finally, we calculate the total value as a function of the threshold, $V(\tau)$, for all models with the reject option at all possible rejection thresholds ($\tau$).
When $\tau\in[0.0, 0.5]$, all predictions are accepted since the confidence of all predictions is above 0.5 in the case of binary classification.
On the other hand, $\tau=1.0$ implies that all predictions are rejected.
We use the $v$ values of the ME scale from~\cref{tab:costs-reliability} to plot the results of all three models in~\cref{fig:metric-values-seen,fig:metric-values-unseen} using~\cref{for:final-V}.
The diamond-shaped markers indicate the optimal confidence thresholds for rejection at which the model achieves the highest total value.

\begin{figure}[h]
\centering
    \setkeys{Gin}{width=\linewidth} %set image parameters
    \begin{subfigure}[b]{.49\linewidth}
        \centering
        \includegraphics[width=\linewidth]{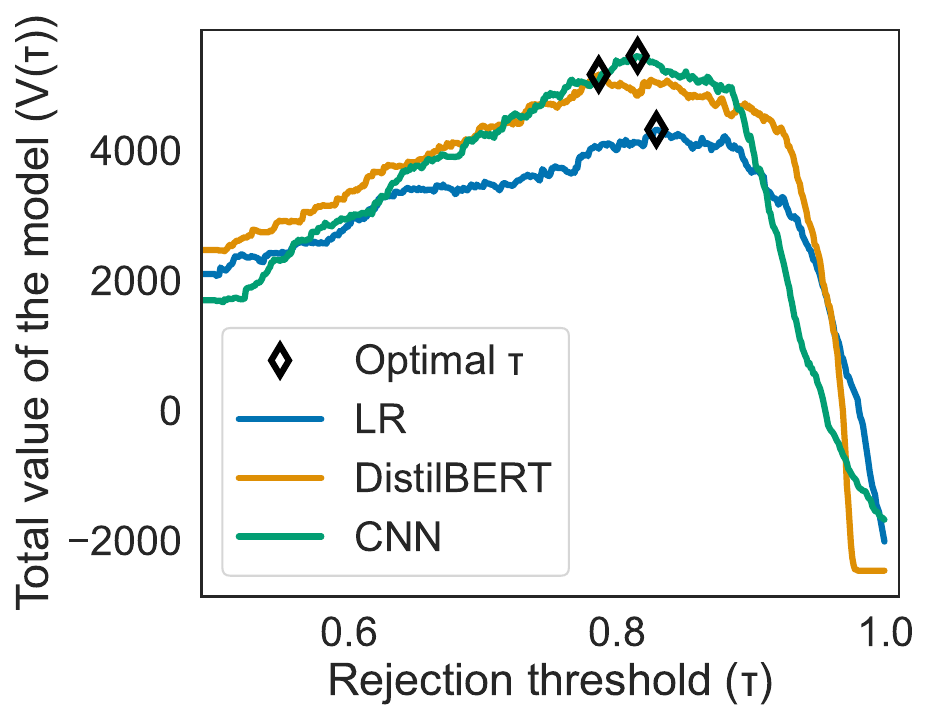}
        \caption{evaluated on \textit{seen} data}
        \label{fig:metric-values-seen-tptn0}
    \end{subfigure}
    \hfill
    \begin{subfigure}[b]{.49\linewidth}
        \centering
        \includegraphics[width=\linewidth]{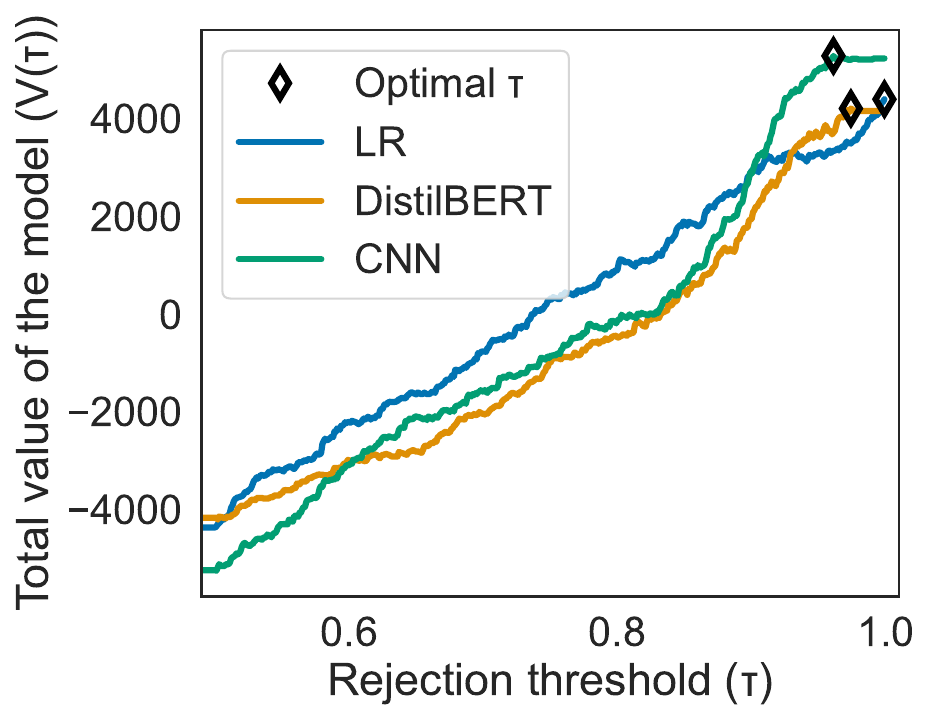}
        \caption{evaluated on \textit{unseen} data}
        \label{fig:metric-values-unseen-tptn0}
    \end{subfigure}
    \label{fig:metric-tptn0}
    \caption{$V(\tau)$ curves of all models with $v$ of TP=0.0, TN=0.0, FP=-16.69, FN=-28.08, and rejection=4.82.}
    \vspace{-0.1in}
\end{figure}

Participants ascribe higher absolute values to TP and TN scenarios compared to FP and FN ones (see~\cref{tab:costs-reliability}), which results in all but one model having the highest value when all predictions are accepted (see~\cref{fig:metric-values-seen,fig:metric-values-unseen}).
%
%This effect can also be observed in~\cref{fig:metric-values-seen,fig:metric-values-unseen}, where for all but one model (LR for \textit{unseen} data) we achieve the highest value when all predictions are accepted.
%
The rejection rates (i.e., the percentage of rejected predictions) and accuracies of accepted predictions at the optimal threshold across the three classifiers can be seen in the first two rows of~\cref{tab:metric}.
\begin{table*}[]
\small
\centering
\begin{tabular}{lcccccccccc}
\toprule
& \multicolumn{3}{c}{\textbf{LR}} & \multicolumn{3}{c}{\textbf{DistilBERT}} & \multicolumn{3}{c}{\textbf{CNN}}\\
\cmidrule(l){2-4} \cmidrule(l){5-7} \cmidrule(l){8-10}
& $\boldsymbol{\tau}$ & \textbf{Acc}  & \textbf{RR} & $\boldsymbol{\tau}$ & \textbf{Acc} & \textbf{RR} & $\boldsymbol{\tau}$ & \textbf{Acc} & \textbf{RR} \\
\midrule
\textbf{Seen data}    & 0.500 & 0.853 & 0.000    & 0.500 & 0.853 & 0.000    & 0.500 &  0.845 & 0.000\\
\textbf{Unseen data}  & 0.531 & 0.646 & 0.043    & 0.500 & 0.643 & 0.000    & 0.500 &  0.624 & 0.000\\
\midrule
\textbf{Seen data ($V_{TP}=0, V_{TN}=0)$}   & 0.829 & 0.925 & 0.316    & 0.786 & 0.923 & 0.202    & 0.815 & 0.934 & 0.299 \\
\textbf{Unseen data ($V_{TP}=0, V_{TN}=0)$} & 0.999 & 0.818 & 0.991    & 0.974 & 1.000   & 0.996    & 0.961 & 0.833 & 0.980 \\
\bottomrule
\end{tabular}
\caption{The optimal rejection thresholds ($\tau$), the accuracy of the accepted predictions (Acc), and rejection rates (RR) of all models for both datasets using the values from the survey.}
\label{tab:metric}
\end{table*}
\begin{table*}[]
\small
\centering
\begin{tabular}{lccccccc}
\toprule
& \multicolumn{2}{c}{\textbf{LR}} & \multicolumn{2}{c}{\textbf{DistilBERT}} & \multicolumn{2}{c}{\textbf{CNN}}\\
\cmidrule(l){2-3} \cmidrule(l){4-5} \cmidrule(l){6-7}
& $\boldsymbol{V(\tau_O)}$ & \textbf{Acc} & $\boldsymbol{V(\tau_O)}$ & \textbf{Acc} & $\boldsymbol{V(\tau_O)}$ & \textbf{Acc} \\
\midrule
\textbf{Seen data}                  & 45534 & 0.853     & 45250 & 0.853    & 44893 & 0.845 \\
\textbf{Unseen data}                & 18563 & 0.631     & 19132 & 0.643    & 18385 & 0.624 \\
\midrule
\textbf{Seen data ($V_{TP}=0, V_{TN}=0)$}     & 4325 & 0.853      & 5172 & 0.853     & 5460 & 0.845 \\
\textbf{Unseen data ($V_{TP}=0, V_{TN}=0)$}   & 4404 & 0.631      & 4213 & 0.643     & 5291  & 0.624\\
\bottomrule
\end{tabular}
\caption{The total values $V(\tau_O)$ and the accuracies (Acc) of all models. Here, $\tau_O$ is the optimal rejection threshold.}
\label{tab:metric2}
\end{table*}
If we were to take the view that the users' baseline expectation is correct machine decisions, then we can set the value of TP and TN to 0.0 and repeat our analysis to examine how $V(\tau)$ behaves as we consider only punishing incorrect predictions without rewarding correct predictions made by the model (considering the regulation effect discussed in~\cref{s:3}).
%To examine how $V(\tau)$ behaves when we only consider the value of FP, FN, and rejected predictions, therefore focusing on punishing incorrect predictions instead of rewarding correct predictions, we set the value of TP and TN equal to 0.0 in~\cref{fig:metric-values-seen-tptn0,fig:metric-values-unseen-tptn0}.
%
Figures~\ref{fig:metric-values-seen-tptn0} and~\ref{fig:metric-values-unseen-tptn0} demonstrate that the optimal values are achieved at increased rejection thresholds ($\tau$).
The last two rows of~\cref{tab:metric} show that the optimal $\tau$ values result in higher accuracies for the \textit{seen} data while rejecting 31.6\% of predictions.
For the \textit{unseen} data, we achieve high accuracies but reject a large fraction of the predictions.

We also compare the effect of using value and the widely-used accuracy metric in selecting the best model, shown in~\cref{tab:metric2}. We observe that both metrics return the same optimal model when correct predictions are rewarded, though there is a difference between \emph{seen} and \emph{unseen} cases. When only incorrect predictions are punished, the optimal models are different as measured by the two metrics: in the case of \emph{seen} data, both LR and DistilBERT perform better than CNN when measured by accuracy, while CNN delivers the highest value; the same observation holds true in the case of \emph{unseen} data -- where the optimal model switches from DistilBERT to CNN when we consider the value they deliver instead of  accuracy.

% xx discuss results xx. Second, we created the Accuracy-Rejection curves of both models evaluated on both datasets and included the optimal rejection threshold points as well in \cref{fig:accuracy-rejection}. xx discuss results, perhaps we can use these plots to show that only focusing on accuracy metrics is not enough. perhaps the plots indicate we can even reject fewer samples and keep high accuracy while the optimal rejection threshold indicates that we should reject more samples because the stakes are too high xx. Finally, in \cref{fig:metric-values}, we present the resulting metric values for all possible rejection thresholds between 0\% (rejecting nothing) and 100\% (rejecting everything). The optimal rejection thresholds are indicated with points on the curves.

% \begin{figure}[h]
%     \centering
%     \includegraphics[width=\columnwidth]{images/accuracy-rejection.png}
%     \caption{Accuracy-Rejection curves for all models evaluated on both the \citet{waseem2016hateful} and the \citet{basile2019semeval} datasets. The points indicate the optimal rejection thresholds resulting from the metric. xx TEST IMAGE REPLACE THIS xx}
%     \label{fig:accuracy-rejection}
% \end{figure}

\section{Discussion}
\label{sec:discussion}

\subsection{Value Ratios, Reliability, and Validity}
%\textbf{value ratio}
Our results show that TP and TN scenarios are highly valued. 
Participants seem to value correct predictions more than incorrect predictions across all scenarios, regardless of whether they are positive or negative.
%
%\color{red}
The value of rejected predictions is the closest to 0 (neutral), as expected, due to them not contributing any benefit or harm, but just delaying the publishing of the post due to the additional human moderation effort.
%\color{black}
%\ych{Maybe explain why this is expected?}
%
For both scales, we observe the same relation of scenarios in terms of values (FN$<$FP$<$Rejection$<$TP$<$TN).
The fact that correct decisions receive higher value ratings indicates strong user appreciation of correct machine decisions. 
%
%\color{red}
The value of FN having a larger magnitude than the value of FP is noteworthy, as users appear to be more negatively affected when a non-hateful post is subject to moderation than when an instance of hate speech is classified as non-hateful. This implies that users would rather contend with an instance of hate speech than have an innocent user punished for a non-hateful post. This phenomenon may be explained by the Blackstone principle from the domain of criminal law: ``Better that ten guilty persons escape, than that one innocent suffer''~\cite{epps2014consequences}. % Our findings This further strengthens our assumption that effective moderation through human-AI collaboration is key to maximizing total value.
% \color{black}
%\ych{Maybe explain why this is expected?}
%
% Participants perceive the impact of incorrectly classifying hateful content as greater than incorrectly classifying non-hateful content.
%We expected that the value of FN would be lower than of FP, so the impact of incorrectly classifying hateful content is considered greater than incorrectly classifying non-hateful content.
%
However, we do consider it surprising that the value of TN is greater than the value of TP.
One possible reason could be that people disagree more on what is considered hateful among the TP scenarios.
We also encountered this phenomenon in the survey results where most people rated TN cases as non-hateful, while for the TP cases there were more disagreements.

%\textbf{reliability}
Regarding reliability, Krippendorff's alpha, $\alpha$, for the 100-level scale being lower than the one for the ME scale is unexpected, as the 100-level scale is bounded with fewer possible options.
The stronger agreement for the ME scale indicates that it is indeed suitable for this task.
Since $\alpha$ compares the expected difference with the observed difference, it follows that the alpha values for the entire scale should be greater than for the individual scenarios.
Generally, participants tend to have low agreement on TP, TN, and rejection cases while they have a high agreement regarding the FP and FN cases. Users tend to agree more regarding what constitutes a misclassified instance than what constitutes a correctly classified instance. For the ME scale, we even observe systematic disagreement for the rejection case, as can be seen by its negative $\alpha$ value. 
This indicates that users are lower in agreement than one would expect by chance, showing the wide variety of opinions regarding rejection cases by users.
%\jie{isn't the result the other way around?} \jie{also, we need to explain what the result entails.}
%
%\color{red}
By considering all answers, instead of answers for certain scenarios, we observe a greatly increased $\alpha$, as the observed difference between ratings is closer to the difference expected by chance. % \jie{what do we mean by expected difference here?}
%
%This does not hold when computing the value for only one scenario.
%
%\color{red}
For example, participants tend to agree on the classification of a single scenario, e.g. TP, but may give different values on both scales, resulting in lower $\alpha$ for the scenario but greater $\alpha$ across all scenarios.
%\color{black}
%\ych{Please check if the red-marked one is correct}
%
Beyond this, the low reliability for the positive compared to negative predictions indicates that participants disagree on what constitutes hate speech in the first place.
    
%\textbf{validity}
Regarding validity, we observe a strong correlation between scales, demonstrating that the ME scale is validated for measuring people's opinions about different hate speech detection scenarios.
The almost S-shaped curve for the data points in~\cref{fig:validity} is due to the lower and upper bounds of the 100-level scale that restrict the participants' choices, making them more likely to assign the lowest or highest value.
Meanwhile, the data points corresponding to the ME scale are skewed towards 0 because of the normalization.
    
\subsection{Value Function for Rejection}

The purpose of the reject option is to reject predictions where the risk of an incorrect prediction is too high.
However, when we use all values obtained from the survey to measure the value function $V(\tau)$, the total value of a model with a reject option is maximized by accepting all predictions.
As shown in~\cref{fig:metric-values-seen,fig:metric-values-unseen}, values are positive at the beginning, decline steadily as the rejection threshold increases, and eventually become negative as more predictions are rejected.
%In~\cref{fig:metric-values-seen,fig:metric-values-unseen} we can see how the value steadily declines and eventually becomes negative as more predictions are rejected. 
%
This observation is not surprising, as the absolute values of correct predictions are greater than the absolute values of incorrect predictions (see~\cref{tab:costs-reliability}).

However, instead of rewarding correct predictions, we believe it is more critical to emphasize penalizing incorrect predictions, as hate speech should be moderated effectively to minimize harm.
%users of a social media platform expect hate speech to be moderated effectively but will be adversely impacted should this not be the case.
%
%\color{red}
To study the effects of this we also analyze the behavior of $V(\tau)$ when users do not experience an increase in value through correct classifications, i.e. TP and TN. To achieve this, we set the scenario values $v$ of TP and TN equal to zero. This results in correct predictions effectively only increasing the total value by the $v$ of rejection when accepted and decreasing when rejected, as can be seen in~\cref{for:final-V}.
%\color{black}
%\ych{This sentence is hard to read and not clear to me. Would you mind rewriting it using an easier-to-understand wording?}
%
The result in~\cref{fig:metric-values-seen-tptn0} shows a steady increase in value before it peaks for each of the three models, eventually falling again and becoming negative as almost all predictions are rejected.
Hence, there is a strong incentive to reject some (but not all) predictions for the \textit{seen} data.
At the points where values are maximized, we found an optimal balance between accepting and rejecting predictions.
Figure~\ref{fig:metric-values-unseen-tptn0} shows that the values continually rise for all three models, only peaking as the rejection threshold approaches 1.
This indicates that the model is very uncertain regarding its predictions for the \textit{unseen} data, which may be expected.
%considering the evaluation on the unseen data.
%
Initially, at the 0.5 rejection threshold, the value is negative as all predictions are accepted.
When the rejection threshold increases, the value rises steadily since too many incorrect predictions are made.
%
%\color{red}
This indicates that the model is not performing well at the task (i.e., high confidence false predictions), and thus the optimal condition to reject most predictions makes the unviable model.
%\color{black}
%\ych{Please check if the red-marked sentence is correct.}
%This indicates that the model is not performing well at the task, as rejecting the vast majority of cases put the model's viability into question.

The results show that by penalizing incorrect predictions without rewarding correct predictions, a significant fraction of the predictions can be accepted from all three models. 
%The results show that by focusing on penalizing incorrect predictions, all three models perform well on seen data since we accept a large fraction of its predictions as a result of high confidence.
%
For unseen data, however, very few predictions from these models can be accepted and the majority are rejected. %with a reject option may still be valuable since they achieve high accuracies when a small fraction of the most confident predictions are accepted.
%
% On the other hand, the risk of making incorrect predictions for all three models for unseen data is higher compared to seen data,
% \color{red}
Such a result confirms the bias in the dataset as also found in previous studies \cite{arango2019hate,grondahl2018all}. The results also show the utility of value as a metric in guiding the decision on when to reject machine predictions. Value utility is further confirmed in the results in \cref{tab:metric2} from our experiment on optimal model selection: the best model selected by value is different compared to using accuracy as the metric.
% \%color{black}
%\ych{Where did we mention this? Maybe say things like in section X}

\subsection{Findings, Implications, and Limitations}
%\textbf{RQ}
Our survey study uncovers several interesting findings. First, social media users are more appreciative of correct decisions made by the platform, with an absolute magnitude higher than the (negative) perception of incorrect decisions. Among the correct decisions, users especially appreciate that non-hateful content is correctly identified and not banned. On the other hand, users show a much higher agreement on the negative value of incorrect decisions than correct ones, indicating a strong consensus over the harm (from both identifying hateful content to be non-hateful, and vice versa). These results indicate that while users appreciate correct decisions, minimizing incorrect decisions remains an important task for social media platforms. On the methodological side, we also believe our proposal of using ME for rating human perception can be particularly relevant for research that aims to tackle social science problems through quantitative approaches, like machine learning. 

By integrating value as a parameter into the human-AI collaboration framework for rejecting machine decisions, we show that value can help guide the decision of when to accept machine decisions to reach the optimal value a model can deliver. By showing how the number of acceptable machine decisions changes when the model is applied to a dataset different from the training data, our results confirm findings from previous research that such datasets are biased and hence the trained models are as well. Our results also show that when considering value as an optimization target, the best model selected can be different compared with using accuracy as the metric. We believe these findings can benefit the research community and industry alike, as they present a novel way of using a value-sensitive reject option to increase the utility of human-AI collaboration across domains.
%
% Regarding our research question, we show that the value of a machine learning model can be maximized using the human-based values derived from a crowdsourced survey study to find the optimal rejection threshold. 
% %Regarding our research question, we show that the value of a machine learning model can be maximized using a reject option which makes use of values derived from a crowdsourced study to find the optimal rejection threshold.
% %
% The rejector is therefore both value-sensitive and smart.
% %
% Whether the rejector is beneficial depends on the underlying confidence in machine predictions and how stakeholders who provide the values for different scenarios view their implications and impacts.
% %The actual usefulness of such a rejector depends on the underlying confidence in predictions by the machine learning model and how the stakeholders that provide the values for different scenarios view their potential implications.
% %
% %Further, we show that a key assumption we make towards the baseline of the user.
% %
% It could be that social media users do not expect the model to perform well in hate speech detection.
% %
% Thus, their perspectives may focus on the positive sides of the model rather than evaluating potential harm, which is the phenomenon that we found in the survey study.
%If the user does not expect the model to perform its task as a default, then their responses will focus on the positives of the model, rather than on an evaluation of potential harm.

%\textbf{Limitations}
Our work is limited to a relatively small sample size (68 subjects per scale). %resulting in a 90\% confidence and 10\% MoE.
We expect the results to be more reliable at a larger sample size.
%
%Nevertheless, having 95\% confidence and 5\% MoE requires 384 participants per scale given the size of our population, which would greatly increase the cost of our study for potentially a marginal gain.
%
Besides, optimal confidence threshold determination relies heavily on empirical data, which may not be available in real applications. An easier way for selecting the optimal threshold would be using well-calibrated models, for which the optimal threshold is only dependent on the human-perceived value. % which means more correct predictions with high confidence and fewer wrong predictions with high confidence. 
Although techniques such as Temperature Scaling can help improve the calibration of existing neural networks or transformer models such as DistilBERT, we still observe that all models are predisposed to producing high-confidence errors.
Finally, due to taking the users' standpoint, we do not fully capture the cost of the moderation team being exposed to hate speech. We leave this as possible future work.

\section{Related Work}
\label{s:2}

\subsection{Hate Speech Detection} 
Online hate speech content refers to ``online messages demeaning people on the basis of their race/ethnicity, gender, national origin, or sexual preference''~\cite{lee2020source}.
Its characterizing features are properties of the target of the language, as compared to other types of online conflictual languages, which are defined by the intention of the author such as cyberbullying or flaming~\cite{burbank1994cross, rayner1997summary}.
%
%It also differs from conflictual languages whose definitions put more emphasis on language style but are not necessarily mutually exclusive, such as abusive, profane, or toxic languages. 
%
A large body of discussion can be found on conflictual languages from social sciences, political science, and computer science~\cite{matias2019preventing,tsesis2001hate,waldron2012harm}.
Hate speech-related research in computer science has identified mismatches between the formalization of hateful content and how people perceive such languages~\cite{balayn2021automatic}.
These mismatches conceptually are further reflected in the technical biases of the machine learning systems used for filtering hateful content.
For instance,~\citet{grondahl2018all} found that F1 scores were reduced by up to 69\% when training a hate speech detection model on one dataset and evaluating it using another dataset from a similar source.
Similarly,~\citet{arango2019hate} found that most research in hate speech detection overestimates the performance of the automated methods due to dataset bias.
In response to these findings, our work aims to explore a human-AI collaborative approach for effective hate speech detection.

\subsection{Human-AI Collaboration and Rejection}
Human-AI collaboration aims to exploit the complementarity between the cognitive ability of humans and the scalability of machines to solve complex tasks at scale~\cite{vaughan2017making, Bansal2021team}. 
%The most common way of utilising human computation for machine learning is perhaps data generation. 
%
%For example,~\citet{deng2009imagenet} and~\citet{heilman2010rating} explored ways of annotating data with crowd workers to tackle computer vision and natural language processing tasks.
%Related work includes early explorations of crowdsourced data annotation for model training, such as ImageNet for vision tasks~\cite{deng2009imagenet}, but also efforts of data annotation in language tasks~\cite{heilman2010rating}.
%
%In human computation for machine learning, research has focused on reducing the annotation effort~\cite{yang2018leveraging}, as well as noises arising from varying levels of worker expertise, skills, motivation, task difficulty, or clarity~\cite{dawid1979maximum,sheng2008get,whitehill2009whose}.
%
%Active learning is also used to reduce the annotation effort or cost~\cite{yan2011active,yang2018leveraging}.
%A focal point of research has been reducing the annotation noises arising from varying levels of worker expertise, skills, and motivation, and task difficulty or clarity~\cite{dawid1979maximum,sheng2008get,whitehill2009whose}, or reducing the annotation effort or costs through active learning approaches \cite{yan2011active,yang2018leveraging}.
%
Some work proposed new ways of collaboration, such as learning crowd vote aggregation models from features of the crowd task~\cite{Kamar_2012_combining} and leveraging crowds to learn features of ML models~\cite{flock_2015,Carlos_pattern}.
Recent work has shifted attention to human involvement in providing interpretations of model decisions and evaluating these interpretations~\cite{kyunglee2018perception, ribeiro2016should}.%, as well as debugging the system or the data~\cite{nushi2017human,yang2019scalpel}.
%
%Work can also be found on other ways of human-AI collaboration from learning crowd vote aggregation models from ``features'' of the crowd task by~\citet{Kamar_2012_combining}, to leveraging crowds to learn features of ML models, as in the pioneering paper by \citet{flock_2015} as well as~\citet{Carlos_pattern}.
%
%The line of work more relevant to ours is hybrid human-AI decision-making that we briefly review as follows.
%
%\subsection{Rejecting Machine Decisions}
%
A notable idea for hybrid human-AI decision-making was recently proposed by~\citet{Law_hearth_cscw18}: humans are involved after a machine decision is observed to have low confidence.
%
%paper below doesn't even discuss that?
%Similar ideas regarding the rejection of machine decisions have been discussed in the past \cite{de2000reject}.
%
Following works can be categorized in several dimensions, namely \emph{when} rejection happens, on \emph{what models}, and based on \emph{what criteria}~\cite{hendrickx2021machine}.
Regarding the ``when'', rejection can be implemented in three ways: the preemptive way where whether a data item needs to be handled by a human is decided beforehand~\cite{coenen2020probability}; the integrated way which uses a rejector inside the machine learning model (e.g., a rejection layer in a neural network) to decide whether a decision should be rejected~\cite{geifman2019reject}; and the dependent way, which is also the most common, which analyzes the rejection option after model decisions~\cite{geifman2017selective, de2000reject, grandvalet2008reject}.
%
%The dependent way is most often used, as we also study in our work.
%
In terms of ``what models'', work has been done on rejecting decisions made by a range of models, such as SVMs~\cite{grandvalet2008reject,coenen2020probability} and different neural networks~\cite{de2000reject,geifman2019reject}.
%
%One should note that in practice neural networks are often poorly calibrated~\cite{Guo2017,Balda2020}, making their decision confidence unreliable.
%An important consideration of rejecting decisions from neural networks is that they are often poorly calibrated~\cite{Guo2017,Balda2020}, making their decision confidence unreliable.
%
In our case, we apply the dependent way to reject models that are based on neural networks.
In terms of ``what criteria'',~\citet{geifman2017selective} proposed a rejection function based on a predefined risk value, an idea also explored in~\cite{nadeem2009reject}.
But unlike ours, their proposals do not consider the impact of machine decisions in a specific context.
The most relevant proposal to our work is from~\citet{de2000reject}, who studied a confidence metric for determining the optimal rejection threshold.
In their work, the threshold is calculated with simulations based on a set of predictions.
Going beyond defining cost values from simulations, our approach determines cost values based on users' perception of machine decisions using a survey study with crowd workers.
%
%Thus, we obtain a threshold that captures the implications of machine decisions from a human perspective.

\subsection{Value Assessment and Measurement}
%Due to algorithmic biases, value of machine learning systems has recently been intensively discussed across the ethics, design, and computer science domains.
%
Value is generally defined as desirable properties of an entity~\cite{birhane2021values}. Specifically for machine learning systems \citet{yurrita2022towards} have identified relevant properties, including individual empowerment, conservation, universalism, and openness.
%
%A big challenge in creating machine learning systems that conform to these properties is the operationalization of value, which means identifying context-specific values and translating them into concrete system specifications~\cite{shahin2021operationalizing}.
%
%Efforts have been made from both social and technological sides.
%
Examples include outlining ethical principles of algorithmic systems~\cite{fjeld2020principled}, developing value-based assessment frameworks~\cite{yurrita2022towards}, and proposing new metrics for evaluating machine learning systems that incorporate value parameters~\cite{casati2021value}. 
%
%Multiple measurement scales, including ME, were compared by \citet{checco2016mag} in a crowdsourced rating setting.
%
However, a research gap in measuring value in social contexts has been identified by~\citet{olteanu2017detection}, who investigated human-centered metrics for machine learning evaluation in hate speech detection.
Their work highlights the gap between accuracy-based evaluation metrics and user perception.
Our work represents a first step towards filling the gap in the context of hate speech detection using ME with a crowdsourced survey.
%
%To do so, we specifically consider and compare three types of measurement scales as part, namely Likert, magnitude estimation \cite{stevens1956direct}, and a 100-level scale \cite{roitero2018fine}, which we discuss in detail in the next sections.

\section{Conclusions}
\label{sec:conclusion}

This paper studies the operationalization and integration of value into human-AI collaboration for hate speech detection. We introduce a value-sensitive rejection mechanism for machine decisions that takes into account the implications of decisions from a user-centered standpoint. We propose 
ME to measure users' value perception regarding different hate speech detection scenarios. To validate ME, we design a survey study, showing that it can provide a reliable, human-centered assessment of the value a machine learning model delivers. Our survey study uncovers a series of interesting findings on user perception. In particular, participants appreciate correct decisions made by the platform, while they show a strong consensus over the harm of incorrect decisions. 
%To demonstrate the utility of value assessment in human-AI collaboration, we experimented with state-of-the-art hate speech detection models on real-world datasets. 
Our results show that value assessment performed by means of ME can guide us to select the best confidence threshold for rejecting machine decisions, thereby maximizing model value and potentially leading to a different best model than when using accuracy.
% The results indicate a discrepancy between the general practice in machine learning assessment and the actual value a model can provide in a specific context. 

% how a value-sensitive smart rejector allows us to take the implications of machine decisions into account when evaluating them.
% %
% We find that Magnitude Estimation can be used for collecting participants' opinions regarding different hate speech detection scenarios while providing reliable results.
% %
% We propose a method for evaluating the value of machine learning models using the values of true positive, true negative, false positive, false negative, and rejected predictions.
% %
% Beyond this, we show that the total value of a machine learning model with a rejector can be formulated by considering users' perspectives towards instances of hate speech. 
%
%\color{red}
% Specifically, when running the model on the test set of a similar source, the rejector shows promising results and the total value for each model used can be maximized through the use of it. 
% %\color{black}
% %\ych{Please check if this red-marked sentence is correct.}
% %
% %Under no circumstances does the rejector reduce total value.
% %
% Our value-sensitive approach, therefore, has the potential to improve user experiences and reduce harm in the context of social media platforms.

%\begin{acks}
%To Robert, for the bagels and explaining CMYK and color spaces.
%\end{acks}

\bibliographystyle{ACM-Reference-Format}
\bibliography{main}


\appendix
\section{Survey}
\label{app:b}
\subsection{Variables}%Independent, Confounding, Control, and Dependent Variables}
The independent variables are the possible scenarios (TP, TN, FP, FN, and rejection).
%
%The true positive (TP) and true negative (TN) scenarios mean that SocialNet successfully detects whether a post is hateful or not, respectively.
%
%The false positive (FP) scenario means that SocialNet incorrectly predicts a non-hateful post as hateful, and conversely for the false negative (FN) scenario.
%
We inform participants in the survey that when hate speech is detected, SocialNet ranks the hateful post lower so that it takes much more effort for the users to find the post.
%
%The rejection scenario means that SocialNet is uncertain whether the post was hateful or non-hateful and thus the post requires human decision-making.
%
For the rejection scenario, we inform the participants in the survey that a moderator needs to check the post within 24 hours, and meanwhile, the post remains visible.% with its original rank on the page.
The design decision of using 24 hours is based on the German NetzDG law, which allows the government to fine social media platforms if they do not remove illegal hate speech within 24 hours~\cite{Tworek2019AnAO}.
%
%\subsection{Confounding Variables}
%Confounding variables are demographics, such as gender, education, location, and age.
%
%Participants from different backgrounds may define hate speech differently and have various perspectives about handling hate speech.
%
%Some previous work focuses on studying confounding variables such as demographics. For example, \citet{gold2018women} showed that there is no significant difference when perceiving hate between genders.
%
%Our research does not study the effect of confounding variables though we may report them to inform the readers about the study context.
%
%We leave the investigation of the effects of demographics on the outcomes to future research.
%\subsection{Control variables}
Our study has two control variables: the measurement scales and the content of posts.
%
%Regarding the content of the post, due to limited resources, we manually select 40 posts from existing datasets containing hateful, neutral, and non-hateful tweets.
%
%The selection procedure is explained in a later subsection.
%
Regarding scales, as described before, we choose ME as our primary scale and use the 100-level scale for validation.
%
%We leave the study of other scales to future work.
%\subsection{Dependent variables}
Our dependent variables are reliability, validity, and value ratios.
We use Krippendorff's alpha to compute reliability, where a value equal to or larger than 0.8 and 0.6 indicates reliable and tentative conclusions, respectively~\cite{krippendorff2004reliability, maddalena2017crowdsourcing}.
Regarding validity, we use convergent validity~\cite{fitzner2007reliability} between the two scales to assess if they measure the same phenomenon.
The value ratio variable describes the perceived value of the scenarios, which is measured by calculating the median of the normalized magnitude estimates of each decision scenario.

\subsection{Demographics}
We perform a demographic analysis regarding the sex, student status, continent of residence, nationality, language, and ethnicity of our participants to study whether there are statistical differences in how they perceive hate speech. We have multiple groups for nationality, ethnicity, and language and two groups for the features of student status, sex, and continent of residence (since only two continents were present in the demographic data of all participants).

\begin{table*}[]
    \small
    \centering
    \begin{tabular}{lccc|ccc}
        \toprule
                     & \multicolumn{3}{c}{\textbf{Two groups}} & \multicolumn{3}{c}{\textbf{More than two groups}}                                                                                                                                                                       \\
        \midrule
                     & \multicolumn{1}{c}{\textbf{Sex}}        & \multicolumn{1}{c}{\textbf{Student}}              & \multicolumn{1}{c}{\textbf{Continent}} & \multicolumn{1}{c}{\textbf{Nationality}} & \multicolumn{1}{c}{\textbf{Language}}  & \multicolumn{1}{c}{\textbf{Ethnicity}} \\
        \midrule
        \textbf{TP}  & 0.302                                   & \textbf{0.032}            & 0.286                                  & 0.218                                    & 0.109                                  & 0.242                                  \\
        \textbf{TN}  & 0.726                                   & 0.379                                             & 0.204                                  & 0.190                                    & 0.216                                  & 0.281                                  \\
        \textbf{FP}  & 0.699                                   & 0.933                                             & 0.073                                  & \textbf{0.020}   & \textbf{0.040} & \textbf{0.037} \\
        \textbf{FN}  & 0.961                                   & 0.150                                             & 0.847                                  & 0.478                                    & 0.438                                  & 0.584                                  \\
        \textbf{REJ} & 0.835                                   & 0.625                                             & 0.496                                  & 0.271                                    & 0.103                                  & 0.068                                  \\
        \bottomrule
    \end{tabular}
    \caption{\textbf{Statistical differences between different groups of participants for various demographic characteristics for each aggregated scenario type in the ME survey. Each cell contains the $p$ value of either the Mann-Whitney U test for two groups or the Kruskal-Wallis test for more than two groups. Bold text indicates significant statistical differences.}}
    \label{tab:results-differences-grp}
\end{table*}

Table \ref{tab:results-differences-grp} shows the group differences for the aggregated scenario types (TP, TN, FP, FN, and REJ) and highlights the $p$ values that differ substantially. We found the most significant differences across all scenarios for different nationalities and languages of participants.

\subsection{Pilot Study}

We conducted a pilot study with 24 participants to test the survey and estimate the required completion time.
The pilot study showed low inter-rater reliability, which may be due to unmatched expectations and unclear scenario descriptions.
Initially, our scenarios mentioned that machine-detected hateful posts would be removed, which could be controversial.
Thus, we changed scenarios to rank the hateful posts lower on the feed. % so that it takes more effort for users to scroll the page to access the posts.
Moreover, we added more lengthy descriptions to explain the potential consequences of all scenarios in a neutral tone, which prepares the participants to focus on evaluating harm (instead of giving rewards).
%
%After the changes, the official survey study shows higher inter-rater reliability.

\subsection{Example Tasks}
\textbf{WARNING: GRAPHIC LANGUAGE.}
\subsubsection{FN scenario with ME scale}
\begin{flushleft}
    Suppose we have the following post:\\
    \textit{``I want this stupid bitch out of my head I've never even liked her lmao I just basically want to rape her''} (from \citet{basile2019semeval})\\
\end{flushleft}
\begin{flushleft}
    \textbf{Question 1}:\\
    Please indicate whether you feel that this post is hateful or not hateful.\\
    **Select input with options: 'hateful' or 'not hateful'**\\
\end{flushleft}
\begin{flushleft}
    \textbf{Question 2}:\\
    You are a user of the SocialNet platform and just saw this post on your main feed because SocialNet's automated detection system is confident that it is not hateful.
\end{flushleft}
\begin{itemize}
    \item This post remains visible on other people's main feeds as well.
    \item If the post is hateful after all, then the detection system was incorrect. This hateful post is now visible on people's main feeds with the consequence that they can get harmed.
    \item If the post is indeed not hateful, then the detection system was correct.
    \item You should be aware that it is expected for SocialNet to correctly classify hate speech. Wrong classifications are undesirable as they may cause harm to people.
\end{itemize}
\begin{flushleft}
    Please indicate whether you agree, disagree, or are neutral about SocialNet's decision.\\
    **Select input with options: 'agree', 'neutral', or 'disagree'.**\\
\end{flushleft}
\begin{flushleft}
    \textbf{Question 3 (only visible when 'agree' or 'disagree' are selected):}\\
    Please indicate how much you agree/disagree with SocialNet's decision using any positive number.\\
    **Show ME input field that allows all positive values greater than 0.**
\end{flushleft}
\subsubsection{FP scenario with 100-level scale}
\begin{flushleft}
    Suppose we have the following post:
    \textit{``Ontario concerns were not listened at immigration meeting in Ottawa For illegal crossing''} (from \citet{basile2019semeval})\\
\end{flushleft}
\begin{flushleft}
    \textbf{Question 1}:\\
    Please indicate whether you feel that this post is hateful or not hateful.\\
    **Select input with options: 'hateful' or 'not hateful'**\\
\end{flushleft}
\begin{flushleft}
    \textbf{Question 2}:\\
    You are a user of the SocialNet platform and have not seen this post on your main feed because SocialNet's automated detection system is confident that it is hateful.
\end{flushleft}
\begin{itemize}
    \item You can still find this post when you scroll down your feed since SocialNet ranks hateful posts lower.
    \item If the post is not hateful after all, then the detection system was incorrect. This neutral post is now ranked lower on people's feeds with the consequence that the post cannot easily reach the author's followers.
    \item If the post is indeed hateful, then the detection system was correct.
    \item You should be aware that it is expected for SocialNet to correctly classify hate speech. Wrong classifications are undesirable as they may cause harm to people.
\end{itemize}
\begin{flushleft}
    Please indicate whether you agree, disagree, or are neutral about SocialNet's decision.
    **Select input with options: 'agree', 'neutral', or 'disagree'.**\\
\end{flushleft}
\begin{flushleft}
    \textbf{Question 3 (only visible when 'agree' or 'disagree' are selected):}\\
    Please indicate how much you agree/disagree with SocialNet's decision using any positive number from 1 to 100. If you feel neutral about SocialNet's decision, select neutral in the field above.
    **Show a numerical slider with values between 1 and 100.**\\
\end{flushleft}
\subsubsection{Rejection scenario with 100-level scale}
\begin{flushleft}
    Suppose we have the following post:
    \textit{``Ever been so hungover that your stomach feels like it's eating itself''} (from \citet{basile2019semeval})\\
\end{flushleft}
\begin{flushleft}
    \textbf{Question 1}:\\
    Please indicate whether you feel that this post is hateful or not hateful.\\
    **Select input with options: 'hateful' or 'not hateful'**\\
\end{flushleft}
\begin{flushleft}
    \textbf{Question 2}:\\
    You are a user of the SocialNet platform and just saw this post on your main feed because SocialNet's automated detection system was not confident enough in whether it was hateful or not.
\end{flushleft}
\begin{itemize}
    \item An internal human moderator at SocialNet needs to look at it within at most 24 hours.
    \item Meanwhile, the post remains visible on people's main feeds.
\end{itemize}
\begin{flushleft}
    Please indicate whether you agree, disagree, or are neutral about SocialNet's decision.\\
    **Select input with options: 'agree', 'neutral', or 'disagree'.**\\
\end{flushleft}
\begin{flushleft}
    \textbf{Question 3 (only visible when 'agree' or 'disagree' are selected):}\\
    Please indicate how much you agree/disagree with SocialNet's decision using any positive number.\\
    **Show a numerical slider with values between 1 and 100.**\\
\end{flushleft}

\end{document}